# Effect of Electro-Diffusion Current Flow on Electrostatic Screening in Aqueous Pores

**Yang Liu[1*], Jon Sauer[2], and Robert W. Dutton[1]**

[1]Center for Integrated Systems, Stanford University, Stanford, CA 94305-4075

[2]Eagle Research & Development, LLC, Boulder, Colorado, 80301

[*]Email: yangliu@gloworm.stanford.edu

*Abstract—* **A numerical study within the framework of the Poisson-Nernst-Planck equations is conducted to investigate electrostatic screening of charged bio-molecules within synthetic pores having diameters of at least 10 Debye lengths. We show that with external biases, the bio-molecule charge is only partially screened due to the presence of electro-diffusion current flow. This is considerably different from the equilibrium Debye-Huckel screening behavior and will result in long-range electrostatic interactions. The potential application to direct bio-molecule charge sensing is also discussed.**

Biological and synthetic pores have recently been intensively explored in bio-molecule translocation studies [1-3]. Translocation dynamics [4] and the role of electrostatics [5] were also numerically investigated using the molecular dynamics approach. Based on the measurement of current blocking signals, those translocation studies were usually limited to extremely small pores at the nanometer scale. On the other hand, pore structures with relatively large size compared to the Debye length, $\Lambda_D$, would tremendously relax the constraints on fabrication and therefore are highly desirable in actual applications. In these structures, an important and unanswered question concerns the role of electrostatic screening in the translocation process. The Debye-Huckel theory would suggest a complete screening of any bio-molecule charge by counter-ions within $\Lambda_D$ (1 nm for 100 mM NaCl that emulates physiological conditions). However, such an equilibrium-based picture does not hold in the presence of ionic current flow, which often is the case in pore-based translocation studies. In fact, in previous work of modeling electrically biased conical pores based on the Debye-Huckel theory, the proper value of the screening length was debated [6]. The major result of this Letter shows that when steady-state ionic current flow is introduced by external electrical biasing across the pores, the charge screening of the translocating bio-molecules is substantially suppressed. This is due to the coupling of non-equilibrium ion transport and electrostatics. The partial screening indicates that electrostatic interactions may play an important role at a distance significantly greater than $\Lambda_D$ in aqueous pores. In particular, the results of this work could be exploited for long-range bio-molecule charge detection [7] at distances much greater than the Debye length that is commonly believed to be the fundamental limit [8,9].

In this scaling study, the ion transport in cylindrical pores is modeled by self-consistently coupled Poisson and Nernst-Planck (PNP) equations. This continuum modeling approach has been firmly established in describing transport of mobile ions under the bulk condition [10]. It reduces to the Poisson-Boltzmann equation [11] in the limit of identically zero fluxes at equilibrium. Recently the PNP theory has been broadly applied to simulate ion transport in open ion channels [12,13]. A comparative study of PNP and Brownian dynamics shows that PNP is generally valid when the pore radius is over $2\Lambda_D$ [14], which is the regime of interest here.

For a 1:1 salt, the PNP equations are given by:

$$-\nabla \cdot (\varepsilon \nabla \psi) = \rho_f + q(n_+ - n_-)$$
$$\partial n_+ / \partial t + \nabla \cdot \vec{f}_+ = U_+ \quad ,$$
$$\partial n_- / \partial t + \nabla \cdot \vec{f}_- = U_-$$

where the subscripts +/− correspond to cation/anion, $q$ is the fundamental charge, $n$ the ion concentration, $\rho_f$ the fixed charge density, $\mu$ the ion mobility, and $\psi$ the electrostatic potential. The terms $\vec{f}_\pm$ are the ionic flux densities and are related to the current densities as $\vec{f}_\pm = \pm \vec{j}_\pm / q$. The terms $U_\pm$ are the net generation rates due to ion binding/release and other chemical processes, which are not considered in this present work. The time derivative terms $\partial n_\pm / \partial t$ also become zero at the steady state condition. In the PNP model, the flux driving force is the gradient of the electrochemical potential $\phi_\pm = \pm\psi + (k_B T / q) \cdot \ln(n_\pm)$, where $k_B T$ is the thermal energy. The flux densities are therefore given by

$$\vec{f}_\pm = -\mu_\pm n_\pm \nabla \phi_\pm = -q D_\pm \nabla n_\pm \mp \mu_\pm q n_\pm \nabla \psi,$$

where the Einstein relation $D = \mu k_B T / q$ is implicitly assumed. The nonlinear PNP equations are solved by a general partial differential equation (PDE) solver, PROPHET [15]. Originally developed for simulating carrier transport in semiconductor devices, the PROPHET simulator has been recently extended to study ion transport in OmpF porin ion channels [16] and orientations of proteins with respect to charged surfaces [17].

We focus on a model system with cylindrical symmetry and vertical height of $Z_S$ as shown in the inset of Fig. 1. The ionic currents flow through a central pore of radius $R_0$ in a solid layer. In sensor applications, the solid layer can be a semiconductor (silicon) substrate of a vertically integrated sensing transistor [7]. It is simply modeled as uniform dielectric impermeable to ions since our focus is the underlying physics of current flow and screening in the ionic solution. Only Poisson's equation is solved inside the solid layer, while the coupled PNP equations are solved in the ionic solution. The interface conditions at the solid/solution interface include continuous electrostatic potential and zero normal flux densities. Electrodes are placed at the top and the bottom boundaries with Dirichlet boundary conditions, $\psi = \mp V_e / 2$, respectively. Ideal reservoirs of ion supply are assumed at both boundaries so that $n_{\pm} = n_0$, where $n_0$ is the bulk concentration under equilibrium and set to 1 mM (corresponding to a classically defined $\Lambda_D$ of about 10 nm). Neumann boundary conditions are used at the outer boundary $r = R_0 + W$ to give vanishing radial components of electric flux and ion fluxes. A small impermeable cylinder with distributed fixed charge density is placed at the center to model a heavy, charged biological macromolecule (DNA or protein). We mainly look at the induced potential change at the solution/solid interface, particularly at point P $(r = R_0, z = Z_S/2)$. The dielectric constants for water, the solid layer, and the macromolecule are set to 80, 3, and 4, respectively. The macromolecule is assigned with a uniform charge density $\rho'_f = 2q$ /nm$^3$, approximately equal to the magnitude of the DNA backbone charge density. The cation and anion mobilities are set to $7.62 \times 10^{-8}$ and $7.92 \times 10^{-8}$ m$^2$/Vs, respectively, which are typical for K$^+$ and Cl$^-$ at low molarities [10].

The effect of external biases is studied for a pore with radius $R_0 = 0.3$ μm (~ 30 Debye lengths). The simulated radial potential profiles in the middle of the pore are shown in Fig. 1. From symmetry considerations, the potential would be zero in the absence of the charged molecule at all biases. With the charged molecule at the center, an exponential decay of the potential is observed at zero bias, exactly reproducing the classical 3-D Debye screening behavior, $\psi \propto 1/r \cdot \exp(-r/\Lambda_D)$. On the other hand, in the case of non-trivial external biases, the potential profiles exhibit a qualitatively different behavior: long decay distances are observed in the radial direction. At point P, the induced potential is in the 1~10 mV range and saturates at higher biases. This signal level is readily detectable with a well-designed electronic structure [18].

The logarithmic magnitude of the induced potential change over the entire simulation domain is shown in Fig. 2(a) and (b) for $V_e = 0$ V and 7 V respectively. The induced potential change is obtained as the difference between simulated potential profiles with and without the charged macromolecule for the same bias condition. In the equilibrium case ($V_e$ =0V), non-trivial potential changes occur only within a few Debye lengths around the macromolecule, in agreement with the Debye screening theory. On the other hand, an appreciable potential change can be seen to spread out to a much longer distance at Ve=7 V. In particular, the potential change is still significant (1 mV or greater) at the vertical interface ($r = R_0$). Such distinctive differences in electrostatic behavior between zero and non-zero biases are not restricted to a particular device structure or dimension. Further simulations for devices with varying pore radius, solid layer thickness and dielectric constants give qualitatively similar behavior.

To understand the observed behavior, one needs to realize that the exponentially decaying potential around the introduced charge as modeled by the Debye-Huckel theory is caused by the detailed balance between ion drift and diffusion processes. However, in the presence of an external bias and induced current flow, the requirement of detailed balance is relaxed. The diffusion component does not completely counter-balance the drift component; the screening of the counter-ions is correspondingly reduced. The essence of such an effect is better illustrated by considering a simple one-dimensional example, where analytical solutions can be obtained with reasonable approximations for three different screening scenarios (Appendix 1). The general potential solution for the partially-screened case is composed of both a rapid exponentially decaying component (Debye screening) and a long range tail (Ohmic behavior) in the presence of non-zero current.

In Fig. 3, the simulated radial component of cation flux density, $f_+^r$, is plotted against vertical position for Ve=7 V. Two cases are simulated with and without the charged macromolecules, respectively. We first note that the two positions at $z = 2$ μm and 2.5 μm correspond to the upper and lower edges of the aperture. Therefore, the peaks (with positive or negative sign) at those two positions correspond to the ion flow entering and leaving the aperture, respectively. For the case without the charged macromolecule, the transition of $f_+^r$ between those two peaks is trivial. In contrast, for the case with the charged macromolecule, significant peaks are observed around the central position ($z = Z_S/2 = 2.25$ μm) where the macromolecule is located. This simulation result reflects a key fact that, although the external bias is applied vertically, the ions flow around the charged macromolecule, resulting in the net fluxes and reduced screening in the radial direction. The impact of the pore radius is examined for devices with different pore sizes. The simulated potential at point $P$ is plotted against the pore radii $R_0$ in Fig. 4. An approximate $1/R_0^2$ dependence is found in this range of simulated voltages and radii. This indicates that the sensing transistor reading will exhibit a $1/R_0$ dependence if we assume fixed gain per unit channel width.

The studied effects are essentially caused by non-equilibrium charge transport; it should be applicable to screening phenomena in general. It is particularly interesting to explore the detection resolution limit for sensor applications based on such a long-range effect. Our further simulations of macromolecules with dipole charge profiles at various axial locations indicate that spatial resolution down to tens of nanometers is practically achievable; it could be useful for resolving important features at that length scale such as DNA copy number variations [19]. Such applications are based on relatively large pores and thereby highly cost-effective. This is in contrast to detection schemes based on translocation current, which relies on demanding fabrication techniques with sub-nanometer precision [20,21].

Y.L. and R.W.D. acknowledge support of the Network for Computational Nanotechnology (NSF EEC-0228390).

**Appendix 1: 1-D Analytical Results**

In the 1-D case, we assume a non-zero potential perturbation $\psi(0)$ at $x = 0$ due to the additional charge. We are interested to solve for $\psi(x), x > 0$. The cation/anion current densities $J_\pm$ are constant in 1-D. An integral-differential form of 1-D PNP is given by:

$$d^2\psi(x)/dx^2 = (\beta J_-)/(\mu_-\varepsilon)\cdot \int_0^x \exp\{\beta[\psi(x)-\psi(y)]\}dy + (\beta J_+)/(\mu_+\varepsilon)\cdot \int_0^x \exp\{\beta[\psi(y)-\psi(x)]\}dy$$
$$+ qn_{-0}/\varepsilon \cdot \exp\{\beta[-\psi_0+\psi(x)]\} - qn_{+0}/\varepsilon \cdot \exp\{\beta[\psi_0-\psi(x)]\}$$
,

where $n_{\pm 0}$ are the ion densities at x=0 and $\beta \equiv q/(k_B T)$. We further assume $J_+ = J_- = J_0$ and $\mu_+ = \mu_- = \mu$, for simplicity and denote $\tilde{n}_{\pm 0} = n_{\pm 0}\exp(\pm\beta\psi_0)$. We then discuss three cases of different screening levels.

I) Equilibrium and fully screened case: $J_0 = 0$. The resultant PB equation has a solution,

$\psi(x) = \psi_0 \exp(-x/\Lambda_D)$, for $|\beta\psi(x)| \ll 1$ and $\psi(\infty) = 0$. Here, we have

$1/\Lambda_D^2 \equiv q\beta/\varepsilon \cdot (\tilde{n}_{-0} + \tilde{n}_{+0})$. This is the result of Debye screening.

II) Linear potential drop and unscreened case: $\psi(x)$ is linear, $\psi(x) = \psi_0 + \psi_0' x$, under condition $J_0 = -q\mu\psi_0' n_{-0} = -q\mu\psi_0' n_{+0}$. The net mobile charge is always zero. This is the case of Ohm's law, where no screening effect is present and only the drift current component exists.

III) A general case of partial screening: assume $|\beta\psi(x)| \ll 1$ and to the first order accuracy one obtains: $d^2\psi(x)/dx^2 = \gamma x + \delta + \psi(x)/\Lambda_D^2$, where $\gamma \equiv 2\beta J_0/(\mu\varepsilon)$ and

$\delta \equiv q/\varepsilon \cdot [\tilde{n}_{-0} - \tilde{n}_{+0}]$. The solution is

$$\psi(x) = (\gamma\Lambda_D^3 + \delta\Lambda_D^2 + \psi_0'\Lambda_D + \psi_0)/2 \cdot \exp(x/\Lambda_D) + (-\gamma\Lambda_D^3 + \delta\Lambda_D^2 - \psi_0'\Lambda_D + \psi_0)/2 \cdot \exp(-x/\Lambda_D)$$
$$-\gamma\Lambda_D^2 x - \delta\Lambda_D^2$$

. If $J_0 \neq 0$, the linear term is superimposed to the exponential terms in the solution, leading to a long range tail in the presence of steady-state current flow.

**Figures:**

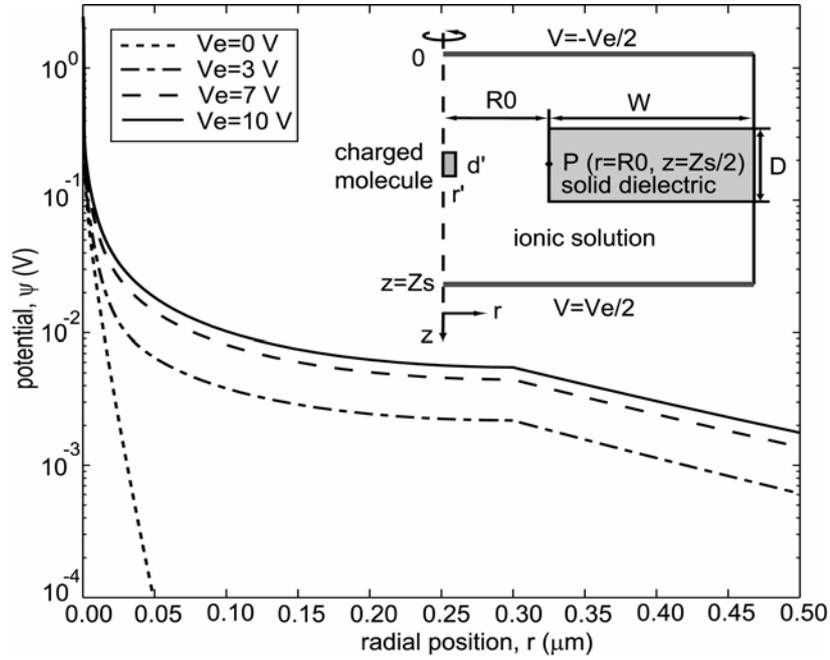

Fig. 1: Simulated electrostatic potential profiles at $z = Z_S/2$ against radial position for different external biases. The aperture radius is 0.3 µm. The inset is a schematic plot of the simulation structure. Cylindrical symmetry is assumed. Default simulation parameters: solid layer thickness D=0.5 µm and width W=2.2 µm, system height $Z_S$ =2.5 µm, macromolecule height $d'$ =20 nm and radius $r'$ =1 nm.

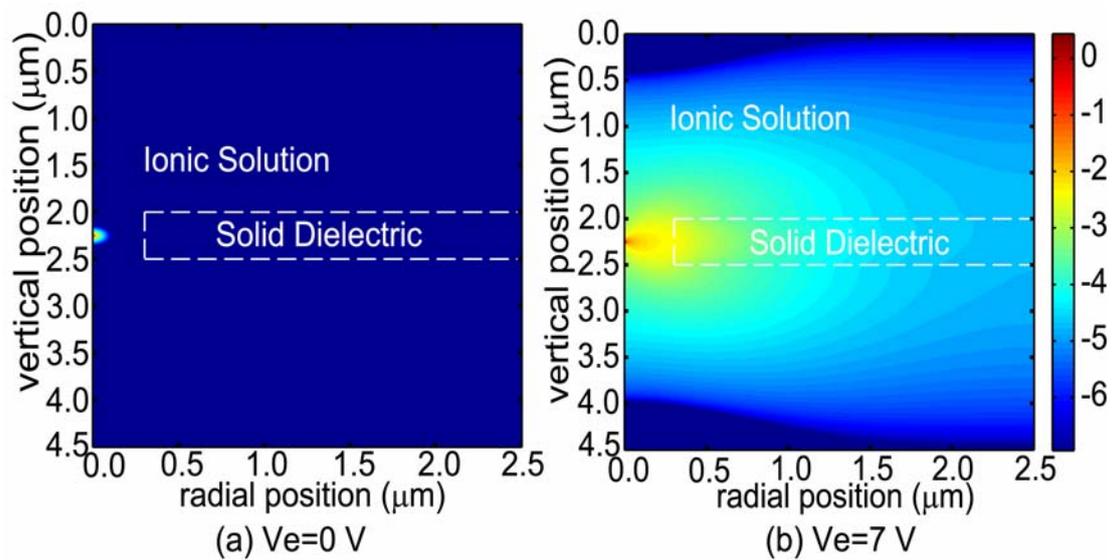

Fig. 2: Simulated 2-D intensity profiles of the potential change induced by the charged macromolecule at the center. (a) $V_e = 0$ V; (b) $V_e = 7$ V. Color scale corresponds to logarithmic magnitude of potential change, $\log_{10}(\Delta\psi)$. The aperture radius is 0.3 μm.

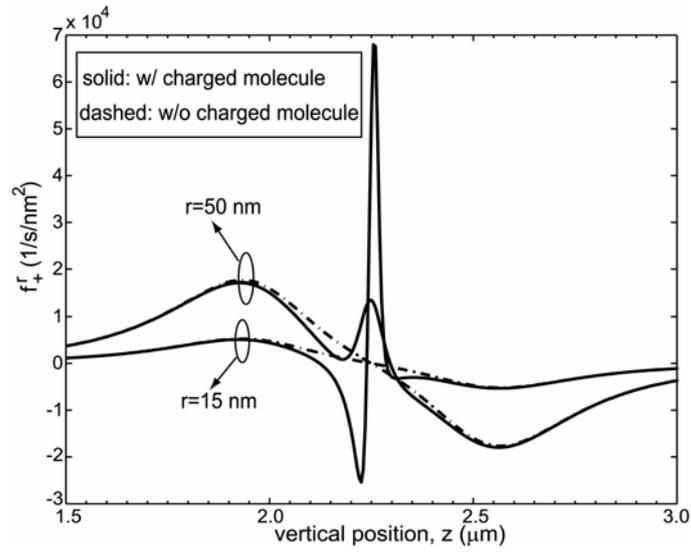

Fig. 3: Simulated radial component of cation flux density, $f_+^r$, at two vertical cut-lines, $r = 15$ nm and 50 nm, against vertical position. Two cases are simulated: with the charged macromolecule (solid) or without (dashed). The aperture radius is 0.3 μm and $V_e = 7$ V.

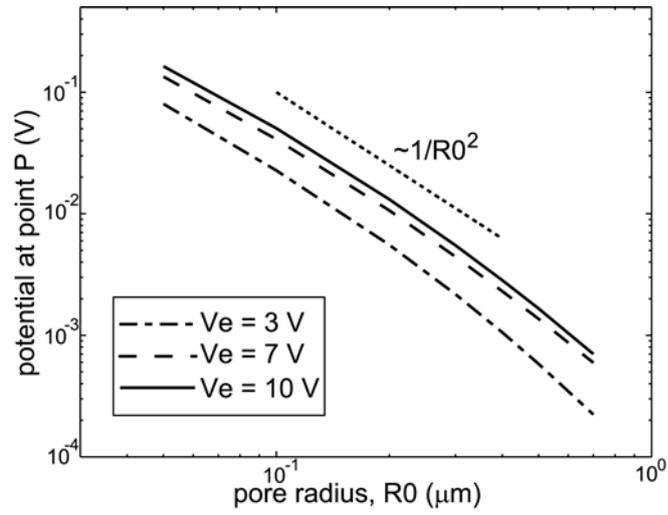

Fig. 4: Dependence of induced potential change at the interface point P (as shown in inset of Fig. 1) on the pore radius for three different biases. A $1/R_0^2$ curve is plotted for visual guide.